\documentclass{article}

\usepackage[english]{babel}
\usepackage{tikz}
\usepackage[letterpaper,top=2cm,bottom=2cm,left=2cm,right=2cm,marginparwidth=1.75cm]{geometry}

\usepackage{amsmath}
\usepackage{amssymb}
\usepackage{graphicx}
\usepackage[colorlinks=true, allcolors=blue]{hyperref}
\usepackage{subcaption}
\usepackage{booktabs}
\usepackage{threeparttable}
\usepackage{diagbox}
\usepackage{pifont}
\usepackage{float}
\usepackage{comment}

\usepackage{mathrsfs} 

\usepackage{adjustbox}

\usepackage{pgfplots}

\usepackage[numbers,sort&compress]{natbib}

\usepackage{siunitx}
\usepackage{physics}
\usepackage{tikz}
\usepackage[outline]{contour} 
\usetikzlibrary{patterns,decorations.pathmorphing}
\usetikzlibrary{arrows.meta}
\tikzset{>=latex}
\contourlength{1.1pt}

\colorlet{mydarkblue}{blue!50!black}
\colorlet{myred}{red!65!black}
\colorlet{vcol}{green!45!black}
\colorlet{watercol}{blue!80!cyan!10!white}
\colorlet{darkwatercol}{blue!80!cyan!20!white}
\colorlet{metalcol}{blue!40!black!10!white}
\tikzstyle{force}=[->,myred,very thick,line cap=round]
\tikzstyle{vvec}=[->,very thick,vcol,line cap=round]
\tikzstyle{piston}=[blue!50!black,top color=blue!30,bottom color=blue!50,middle color=blue!20,shading angle=0]
\tikzstyle{water}=[draw=mydarkblue,top color=watercol!90,bottom color=watercol!90!black,shading angle=5]
\tikzstyle{vertical water}=[water,
  top color=watercol!90!black!90,bottom color=watercol!90!black!90,middle color=watercol!80,shading angle=90]
\tikzstyle{dark water}=[draw=mydarkblue,top color=darkwatercol,bottom color=darkwatercol!80!black,shading angle=5]
\tikzstyle{metal}=[draw=metalcol!20!black,top color=metalcol,bottom color=metalcol!90!black,shading angle=10]
\tikzstyle{width}=[{Latex[length=3,width=3]}-{Latex[length=3,width=3]}]

\newcommand{\Eq}[1]{Eq.~(\ref{#1})}

\usepackage{hyperref}

\title{A scalability Benchmark Study of Model Order Reduction Techniques for very large, strongly coupled
Vibroacoustic Problems} 
\author{Sander Metting van Rijn$^{a\dag}$, Linus Taenzer$^{a}$$^{b}$$^{*\dag}$, Paolo Tiso$^{b}$, Bart Van Damme$^{a}$ \\\\
        \small $^{a}$ Laboratory of Acoustics/ Noise Control, Empa, Überlandstrasse 129, 8600 Dübendorf, Switzerland \\
        \small $^{b}$ Institute for Mechanical Systems, ETH Zürich, Leonhardstraße 21, 8092 Zürich, Switzerland \\
        \small $^{*}$Corresponding author: Linus Taenzer; \tt{linus.taenzer@empa.ch} \\
        \small $^\dag$ These authors contributed equally to this work. \\
}

\date{}

\begin{document}
\maketitle

\begin{abstract}
Model Order Reduction (MOR) can significantly reduce the computational cost of vibroacoustic simulations. 
While most MOR research focuses on single-domain systems (e.g., structural dynamics or computational fluid mechanics), this work compares MOR techniques for large multi-domain problems to identify methods that remain efficient and accurate at very large scales.
In particular, harmonic response simulations of vibroacoustic fluid-structure coupled systems used to compute transfer functions from an input force to either structural acceleration or pressure in the heavy fluid domain are of high interest. To achieve this, the most common MOR techniques based on modal methods and Krylov subspace methods are compared for multi-material systems.\\
To assess the feasibility and accuracy of these techniques for different system sizes, a scalable benchmark model of a water-filled Plexiglass cylinder is developed, with mesh sizes from 10,000 to 1,000,000 Degrees of Freedom (DOF). The quality of the models is assured by validation against experimental data. The geometry, model data, and experimental results are made available so that they can be used as a benchmark for further studies.\\
For systems larger than 100,000 DOF, the investigated modal methods become impractical due to memory limitations, even on powerful workstations. Among the tested techniques, a Krylov subspace two-level orthogonal Arnoldi reduction, combined with symmetrization and conditioning of the system matrices, provides the most accurate and efficient approximation of the target transfer functions—particularly for large-scale models up to 1,000,000 DOFs. This approach achieves a speedup of up to 600× compared to the full model.
\end{abstract}

\section{Introduction}
Numerical modeling of dynamic Fluid Structure Interaction (FSI) problems is a widely used technique in biomechanics and the automotive, maritime, and aerospace industries~\cite{Kim2014, PANDA20161167,Hermann2011,HERRMANN2010153,MERZ2009266,VanDeWalle2017,Maess2007}. Its aim is to predict the acoustic pressure in the fluid domain, resulting from the dynamic deformation of an elastic medium. Generally, a distinction is made between one-way FSI, where the light fluid does not result in a feedback to the structural vibration due to its low impedance, and high-impedance heavy fluids, where a so-called strong coupling (two-way FSI) is required. Strong coupling typically occurs in applications where thin or light components interact with fluids such as water, oil, braking fluids, or partially submerged ship hulls~\cite{Richter2017,Anitie2024}. In many use cases of vibroacoustics, the mean flow is negligible (i.e., the Mach number is much smaller than one), which justifies using the linear wave equation for a stationary fluid to characterize the acoustic field. 

To accurately predict the hydro- and vibro-acoustic behavior of these practical applications in the acoustic frequency range of $20$-$20{,}000$ Hz, three-dimensional models that fully couple the fluid and structural domains are required.
The computational modeling is usually done by either the Finite Element Method (FEM) or the boundary element method~\cite{AtallaSgard2015,Marburg2008,HambricSungNefske2016}. This work focuses on interior acoustic problems, where a vibrating structure surrounds a confined acoustic domain, for which FEM is a common choice~\cite{Fischer2004}. 
As structural models become increasingly complex, for instance for 3D printed and topologically optimized structures, the associated computational cost can grow prohibitively high -- especially in dynamic simulations requiring a matrix factorization for each evaluated frequency step. To mitigate this, Model Order Reduction (MOR) techniques are employed to approximate high-dimensional systems within lower-dimensional subspaces, significantly reducing computational effort while preserving essential system behavior~\cite{Marburg2024FrequencySweeps}. Most MOR methods are projection-based, and the construction of the reduced basis varies across approaches. Substructuring approaches, such as Guyan condensation~\cite{GUYAN1965} and Craig-Bampton~\cite{CraigBampton1968Coupling}, exploit either static condensation or component mode synthesis. Data-driven methods, for example Proper Orthogonal Decomposition (POD)~\cite{LIANG2002P1,LIANG2002P2}, derive the basis from measurement or simulation snapshots. Modal approaches retain only a truncated set of eigenmodes~\cite{ROULEAU2017110,Nefske,Wolf}, while moment-matching methods build Krylov subspaces to capture the system’s input–output behavior~\cite{Su,Bai4,Grimme}. Finally, control-oriented approaches such as balanced truncation~\cite{Moore} construct bases from controllability and observability information. Each method, however, has specific drawbacks:
Guyan condensation requires time-intensive corrections for dynamic problems, and FE models with large interface regions limit the practicality of substructuring techniques such as the Craig–Bampton method due to expensive matrix factorizations. Generating sufficient snapshots for AI-based MOR schemes is impractical for very large systems. These approaches are therefore omitted from this study.
Instead, this study focuses on modal and Krylov based projection methods for large scale vibroacoustic FSI systems assembled with a commercial FEM preprocessor. 

Earlier formulations of modal bases include uncoupled structural and fluid bases~\cite{Ben,Maess}, weakly or strongly coupled variants~\cite{Ben,Kim}, and residual‑based corrections~\cite{Maess,Bobillot}. Convergence issues with such schemes have motivated the development of fully coupled modal bases with different symmetrization strategies~\cite{Tabak_IRCA,Tabak_sym}, although their computational cost for large‑scale systems has not yet been explored. The investigated models range from 2D models with several hundred Degrees of Freedom (DOF) to 3D models with $1,000$ to $15,000$ DOF. 
Krylov subspace techniques provide an efficient alternative for forced vibration problems, particularly when rational Krylov subspaces are used~\cite{Brenner2021}. For symmetric dynamical systems, the Lanczos algorithm is standard~\cite{lanczos1950iteration,Wu2000}, while for non-symmetric problems with typical FSI matrices, the Arnoldi algorithm is a popular choice~\cite{Arnoldi1951}. Alternatively, the potential formulation allows the system to be reformulated as a symmetric problem, where the coupling becomes part of the global damping matrix~\cite{Everstine}. 
For modal and Krylov-subspace MOR methods in the presence of FSI, two major challenges arise:
(1) strong differences in magnitude between displacement and pressure DOFs lead to ill‑conditioning, affecting both solvers' efficiency and modal projections accuracy; (2) strong coupling produces non‑symmetric system matrices, complicating Krylov and eigenvalue computations. Scaling strategies have been proposed to mitigate the first issue~\cite{Maess}, while the fluid-potential formulation addresses the second~\cite{Everstine}.
Although modal and Krylov based projection methods are well established, there is limited knowledge about their \textbf{scalability}, \textbf{numerical stability}, and \textbf{practical performance} for systems with up to one million degrees of freedom.
To address this gap, this paper provides:
\begin{itemize}
\item 	a scalable open source benchmark model for systematic MOR studies,
\item 	a comparative evaluation of modal and Krylov methods for systems up to $1,000,000$ DOFs,
\item 	an assessment of symmetrization and preconditioning strategies, and
\item 	a demonstration that Krylov subspace reduction is the only method exhibiting true scalability in the tested setting.
\end{itemize}
With a comprehensive description of all steps of this analysis for large-scale systems, we aim to provide a hands-on best-practice guide for potential users. Moreover, since many studies in this field are purely numerical, we present a finite element model that is experimentally validated against measurements, requiring careful implementation of model complications such as boundary conditions and material damping. This results in a highly realistic, scalable benchmark model that can serve as a quantitative reference for future research.
The paper is structured as follows. Section~\ref{sec:modeling} provides an overview of existing projection‑based methods, including modal and Krylov approaches. We then introduce the benchmark model, which will be used to compare different MOR techniques, and present the experimental setup in Section~\ref{sec:water-filled_cylinder}. The previously mentioned projection methods are compared for the use case of computing transfer functions of displacements and pressures in a water-filled cylinder under harmonic force loading, as detailed in Section~\ref{sec:results}. 

\section{Model order reduction approaches for dynamic systems with fluid structure interaction}
\label{sec:modeling}

\subsection{Description of the full dynamic system}
The general form of the dynamic equation of motion for a linear structural system solved by the FEM can be expressed as
\begin{equation}
    \mathbf{M}  \mathbf{\ddot x}(t) + \mathbf{D}  \mathbf{\dot x}(t) + \mathbf{K} \mathbf{x}(t)= \mathbf{f}(t)
    \label{eq:Gov_Eq}
\end{equation}
where $\mathbf{M}$ is the mass matrix, $\mathbf{D}$ the damping matrix, $\mathbf{K}$ the stiffness matrix, $\mathbf{x}(t)$ the degree of freedom vector and $\mathbf{f}(t)$ the external force vector applied to the system.
To analyze this system in the frequency domain the Laplace transform is applied by using $\mathbf{x}(t) = \hat{\mathbf{x}} e^{s t}$ and a corresponding forcing function $\mathbf{f}(t)=\hat{\mathbf{f}} e^{s t}$. The harmonic response of the dynamic system yields
\begin{equation}
    \left(s^2 \mathbf{M}  + s \mathbf{D}  + \mathbf{K} \right) \hat{\mathbf{x}} = \hat{\mathbf{f}},
\end{equation}
where in steady state $s=i\omega$, $\omega$ being the angular frequency.
The frequency domain formulation is particularly useful when solving for an unknown state variable $\hat{\mathbf{x}}$ for a given force $\hat{\mathbf{f}}$, since only an algebraic equation must be solved instead of a differential equation.
For FSI problems the dynamic system can be formulated either by an unsymmetric formulation based on the coupling between displacement $\mathbf{u}$ and pressure $\mathbf{p}$ or the symmetric potential formulation coupling the fluid potential $\mathbf{v}$ with the displacement $\mathbf{u}$.

\subsubsection{Unsymmetric formulation}
For FSI problems, the governing equations must be adapted to account for both the structural and fluid domains. By using a block matrix approach, and by performing a Laplace transform of the general equation, the coupled system of equations can be expressed as

\begin{gather}
    \Bigg(\begin{bmatrix} \mathbf{K}_s & \mathbf{C}_{fs} \\ \mathbf{0} & \mathbf{K}_f \end{bmatrix} + s \begin{bmatrix} \mathbf{D}_{s} & \mathbf{0} \\ \mathbf{0} & \mathbf{D}_{f} \end{bmatrix} +s^2 \begin{bmatrix} \mathbf{M}_s & \mathbf{0} \\ \mathbf{C}_{sf} & \mathbf{M}_f \end{bmatrix}
    \Bigg)
    \begin{bmatrix} 
    \mathbf{u} \\ \mathbf{p}
    \end{bmatrix} 
    = 
    \begin{bmatrix} 
    \mathbf{f}_s \\ \mathbf{f}_f
    \end{bmatrix}.
\label{eq:gov_eq}
\end{gather}
where $\mathbf{M}_s$, $\mathbf{M}_{f}$, $\mathbf{D}_{s}$, $\mathbf{D}_{f}$, $\mathbf{K}_s$, $\mathbf{K}_f$ are the mass, damping and stiffness matrices of the structural and fluid domains. $\mathbf{C}_{sf}$ and $\mathbf{C}_{fs}$ are the coupling matrices to ensure FSI. $\mathbf{f}_s$ and $\mathbf{f}_f$ are the force vectors acting on the structure and fluid and $\mathbf{u}$ and $\mathbf{p}$ are the displacement and pressure vectors of the solved system. 

\subsubsection{Symmetric formulation}
As explained by~\citet{Everstine}, the computational efficiency of solving a general FSI equation with unsymmetric system matrices can be significantly improved by symmetrizing the mass and stiffness matrices. To achieve this, the formulation is changed from the previously introduced deformation-pressure $\mathbf{u}-\mathbf{p}$ formulation to the displacement-potential $u-\Phi$ formulation.\\
In this case, the state of the fluid is described by the scalar fluid field potential $\boldsymbol{\Phi}$ defined as
\begin{gather}
    \mathbf{p} = -\rho \dot{\boldsymbol{\Phi}} = -\rho s \boldsymbol{\Phi}.
\end{gather}
By using 
\begin{equation}
    \mathbf{f}_{\Phi} := \frac{\mathbf{f}_{f}}{s} \qquad \Rightarrow \quad \mathbf{f}_{f} = \dot{\mathbf{f}}_{\Phi},
\label{eq:sf_Krylov_sym_7}
\end{equation}
the potential formulation can be written as
\begin{gather}
\label{eq:symm}
    \Bigg(\begin{bmatrix} \mathbf{K}_s & \mathbf{0} \\ \mathbf{0} & -\rho\mathbf{K}_f \end{bmatrix} + s \begin{bmatrix} \mathbf{D}_{s} & -\rho\mathbf{C}_{fs} \\ \mathbf{C}_{sf} & -\rho\mathbf{D}_{f} \end{bmatrix} +s^2 \begin{bmatrix} \mathbf{M}_s & \mathbf{0} \\ \mathbf{0} & -\rho\mathbf{M}_f \end{bmatrix}
    \Bigg)
    \begin{bmatrix} 
    \mathbf{u} \\ \boldsymbol{\Phi}
    \end{bmatrix} 
    = 
    \begin{bmatrix} 
    \mathbf{f}_s \\ \mathbf{f}_{\Phi}
    \end{bmatrix}.
\end{gather}
Since $\mathbf{C}_{sf} = -\rho\mathbf{C}_{fs}^{T}$, it follows that all system matrices $\mathbf{M}$, $\mathbf{D}$ and $\mathbf{K}$ are now symmetric.

\subsubsection{Matrix conditioning}
To further enhance both computational efficiency, accuracy, and convergence when
solving the FSI equation, the Frobenius norms of the submatrices can be adjusted to bring their magnitudes into a similar range. This results in a well-conditioned problem as introduced by \citet{Maess}. 
The symmetrized Laplace transform formulation from \Eq{eq:symm} is homogenized and multiplied by conditioning matrices written as
\begin{gather}
\begin{split}
    \begin{bmatrix} b_{1}\mathbf{I} & 0 \\ 0 & b_{2}\mathbf{I} \end{bmatrix} \Bigg(\begin{bmatrix} \mathbf{K}_s & \mathbf{0} \\ \mathbf{0} & -\rho\mathbf{K}_f \end{bmatrix} + s \begin{bmatrix} \mathbf{D}_{s} & -\rho\mathbf{C}_{fs} \\ \mathbf{C}_{sf} & -\rho\mathbf{D}_{f} \end{bmatrix} +s^2 \begin{bmatrix} \mathbf{M}_s & \mathbf{0} \\ \mathbf{0} & -\rho\mathbf{M}_f \end{bmatrix}
    \Bigg) ... \\ ... \begin{bmatrix} a_{1}\mathbf{I} & 0 \\ 0 & a_{2}\mathbf{I} \end{bmatrix}
    \begin{bmatrix} 
    \mathbf{u} \\ \boldsymbol{\Phi}
    \end{bmatrix}
     = 
    \begin{bmatrix} a_{1}b_{1}\mathbf{I} & 0 \\ 0 & a_{2}b_{2}\mathbf{I} \end{bmatrix}
    \begin{bmatrix} 
    \mathbf{f}_s \\ \mathbf{f}_{\Phi}
    \end{bmatrix}.
\end{split}
\end{gather}
Performing the matrix multiplications results in new system matrices that include the coefficients $a_{1}$, $a_{2}$, $b_{1}$ and $b_{2}$
\begin{gather}
\begin{split}
     \mathbf{K} &= \begin{bmatrix} a_{1}b_{1}\mathbf{K}_s & \mathbf{0} \\ \mathbf{0} & -a_{2}b_{2}\rho\mathbf{K}_f \end{bmatrix} \\
     \mathbf{C} &= \begin{bmatrix} a_{1}b_{1}\mathbf{D}_{s} & -a_{2}b_{1}\rho\mathbf{C}_{fs} \\ a_{1}b_{2}\mathbf{C}_{sf} & -a_{2}b_{2}\rho\mathbf{D}_{f} \end{bmatrix} \\
     \mathbf{M} &= \begin{bmatrix} a_{1}b_{1}\mathbf{M}_s & \mathbf{0} \\ \mathbf{0} & -a_{2}b_{2}\rho\mathbf{M}_f \end{bmatrix}.
\end{split}
\end{gather}
Within each system matrix, the Frobenius norms of the submatrices should be of the same magnitude, resulting in the following set of equations

\begin{equation}
\begin{aligned}
     \big\|a_{1}b_{1}\mathbf{M}_{s}\big\| &= \big\|a_{2}b_{2}\rho\mathbf{M}_{f}\big\|,  \\
     \big\|a_{2}b_{1}\rho\mathbf{C}_{fs}\big\| &= \big\|a_{1}b_{2}\mathbf{C}_{sf}\big\|, \\
     \big\|a_{1}b_{1}\mathbf{K}_{s}\big\| &= \big\|a_{2}b_{2}\rho\mathbf{K}_{f}\big\|, \\
     \big\|a_{1}b_{1}\mathbf{D}_{s}\big\| &= \big\|a_{2}b_{2}\rho\mathbf{D}_{f}\big\|. 
\end{aligned}
\end{equation}

Ignoring the last equation, since $\mathbf{D}_{f}$ is typically equal to zero due to negligible viscosity, there remain three equations for the four unknown coefficients. This underdetermined system has infinitely many solutions, so the coefficients $a_{1}$ and $b_{1}$ are set to 1. In the now overdetermined system of equations, the two solutions for each of $a_{2}$ and $b_{2}$ are denoted as $a_{2,1}$, $a_{2,2}$, $b_{1,1}$ and $b_{1,2}$. The final coefficients are obtained by taking the geometric mean of the two possible values for $a_{2}$ and $b_{2}$ written as
\begin{equation}
    a_{2} = \sqrt{a_{2,1}\cdot a_{2,2}} \qquad b_{2} = \sqrt{b_{2,1}\cdot b_{2,2}}.
\end{equation}
Multiplying the found coefficients with the submatrices of the system matrices results in a well-conditioned problem.

\subsection{Projection-based reduced order modeling for FSI problems}
In projection-based MOR, the state-variables $\mathbf{x} \in \mathbb{C}^{n \times 1}$ of the original model are approximated by $\mathbf{\hat{x}} \in \mathbb{C}^{n \times 1}$, which lies in the subspace spanned by the columns of the projection matrix $\mathbf{V} \in \mathbb{C}^{n \times k}$. In this projected subspace, $\mathbf{\hat{x}}$ can now be calculated as a linear combination of the column vectors of $\mathbf{V}$ as 

\begin{equation}
	\mathbf{x} \approx \mathbf{\hat{x}} = \mathbf{V} \mathbf{x}_{r}.
\end{equation}
The weights for the linear combination are contained in $\mathbf{x}_{r} \in \mathbb{C}^{k \times 1}$, which can be interpreted as the reduced state variables.
Through this projection the problem is simplified. It is no longer necessary to calculate all the $n$ unknown state variables of the original model, but only the $k$ state variables of the reduced system.\\
A one-sided projection-based MOR can be applied directly on the symmetric system matrices of a second-order system
\begin{equation}
    \mathbf{M}_{r}  \mathbf{\ddot x}_{r} + \mathbf{D}_{r}  \mathbf{\dot x}_{r} + \mathbf{K}_{r} \mathbf{x}_{r}= \mathbf{f}_{r}
\end{equation}
where $\mathbf{M}_{r} = \mathbf{V}^{H} \mathbf{M} \mathbf{V}$, $\mathbf{D}_{r} = \mathbf{V}^{H} \mathbf{D} \mathbf{V}$, $\mathbf{K}_{r} = \mathbf{V}^{H} \mathbf{K} \mathbf{V}$, and $\mathbf{f}_{r} = \mathbf{V}^{H}\mathbf{f}$. The goal of the wide range of MOR techniques is to efficiently compute a suitable projection basis $\mathbf{V}$. The two most used approaches are presented in the following sections.

\subsubsection{Modal methods}

\paragraph{Uncoupled bases}
As explained by~\citet{Ben}, a straightforward approach to finding a projection basis $\mathbf{V}$ for the FSI problem is to determine a subset of eigenmodes separately within the fluid and structure domains ignoring any coupling terms written as
\begin{equation}
\begin{aligned}
    (\mathbf{K}_{s} + s^2 \mathbf{M}_{s}) \mathbf{u} = \mathbf{0} \\
    (\mathbf{K}_{f} + s^2 \mathbf{M}_{f}) \mathbf{p} = \mathbf{0}.
\end{aligned}
\end{equation}
The subset of $k_{s}$ eigenvectors $\mathbf{v}_{s,i}$ found in the structural domain is used to create the projection matrix $\mathbf{V}_{s} = [\mathbf{v}_{s,1}\, \mathbf{v}_{s,2}\, \ldots \,\mathbf{v}_{s,k_{s}}]$, whereas the $k_{f}$ eigenvectors $\mathbf{v}_{f,i}$ of the fluid domain are used to create the projection matrix $\mathbf{V}_{f} = [\mathbf{v}_{f,1}\, \mathbf{v}_{f,2}\, \ldots \,\mathbf{v}_{f,k_{f}}]$.
The full projection matrix $\mathbf{V}$ can now be assembled using $\mathbf{V}_{s}$ and $\mathbf{V}_{f}$ as
\begin{equation}
\label{eq:projection_matrix}
    \mathbf{V} = \begin{bmatrix} \mathbf{V}_s & \mathbf{0} \\ \mathbf{0} & \mathbf{V}_f \end{bmatrix}.
\end{equation}
In general, for effective MOR, it is important that $k_s+k_f \ll n$, $n$ being the size of the full system, so that the reduced system matrices remain small. The reduction basis is formed by the eigenvectors associated with the $k$ smallest eigenvalues (i.e., the $k$ lowest modes), computed with MATLAB \textit{eigs}, which uses an implicitly restarted Krylov-subspace method (ARPACK / Krylov–Schur)~\cite{LehoucqSorensenYang1998ARPACK,Stewart2001KrylovSchur}.

\paragraph{Weakly Coupled Bases}
This method partly incorporates the coupling between the fluid domain and the structural domain~\cite{Kim}.
Ignoring the dynamic part of \Eq{eq:gov_eq}, the remaining static part can be written as

\begin{equation}
    \begin{bmatrix} \mathbf{K}_s & \mathbf{C}_{fs} \\ \mathbf{0} & \mathbf{K}_f \end{bmatrix} \begin{bmatrix} \mathbf{u} \\ \mathbf{p}\end{bmatrix} = \begin{bmatrix} \mathbf{f}_s \\ \mathbf{f}_f\end{bmatrix}.
\end{equation}
By applying the Schur complement, a relation between the displacement in the structure part $\mathbf{u}$ and the pressure in the fluid domain $\mathbf{p}$ can be found
\begin{equation}
    \mathbf{K}_{s}\mathbf{u} = \mathbf{f}_{s} - \mathbf{C}_{fs} \mathbf{p}.
\end{equation}
Assuming no external forces act on the structure ($\mathbf{f}_{s}=\mathbf{0}$), and by taking the inverse of $\mathbf{K}_{s}$, this yields
\begin{equation}
\begin{split}
    \mathbf{u} =  - \mathbf{K}_{s}^{-1}\mathbf{C}_{fs} \mathbf{p} = \mathbf{P} \mathbf{p},
\end{split}  \label{eq:schur_complement_2} 
\end{equation}
where $\mathbf{P} = - \mathbf{K}_{s}^{-1}\mathbf{C}_{fs}$. 

In applying the uncoupled bases MOR, the displacement vector $\mathbf{u}$ can be approximated by $\hat{\mathbf{u}} = \mathbf{V}_{s} \mathbf{u}_{r}$, and the pressure vector $\mathbf{p}$ by $\hat{\mathbf{p}} = \mathbf{V}_{f} \mathbf{p}_{r}$, where the subscript $r$ denotes again the reduced space. 
Using \Eq{eq:schur_complement_2}, the approximate displacement corrections, stemming from the FSI, can be taken into account when expressing the approximation of the displacement vector $\mathbf{u}$~\cite{Kim}

\begin{equation}
\begin{split}
    \mathbf{u} \approx  \hat{\mathbf{u}} &= \mathbf{V}_{s}\mathbf{u}_{r} + \mathbf{P}\hat{\mathbf{p}}\\
    \mathbf{u} \approx  \hat{\mathbf{u}} &= \mathbf{V}_{s}\mathbf{u}_{r} + \mathbf{P}\mathbf{V}_{f}\mathbf{p}_{r}.
\end{split}   
\label{eq:displacement_approx}
\end{equation}
From this equation, it becomes evident that the projection matrix $\mathbf{V}$ can be expressed as
\begin{equation}
\label{eq:PV_f_decomposition}
    \mathbf{V} = \begin{bmatrix} \mathbf{V}_s & \mathbf{P}\mathbf{V}_{f} \\ \mathbf{0} & \mathbf{V}_f \end{bmatrix}.
\end{equation}
leading to only correction for the displacement field. 

\paragraph{Strongly coupled bases}
Introduced by~\citet{Kim}, this method takes advantage of the fact that the projection matrix $\mathbf{V}$ used in the weakly coupled bases approach can be expressed as a combination of two separate projections. These projections individually account for the characteristics of the structural and fluid domains
\begin{equation}
    \mathbf{V} = \mathbf{V}_{sol} \mathbf{V}_{flu},\quad \mathbf{V}_{sol} = \begin{bmatrix} \mathbf{V}_s & \mathbf{P} \\ \mathbf{0} & \mathbf{I} \end{bmatrix}, \mathbf{V}_{flu} = \begin{bmatrix} \mathbf{I} & \mathbf{0} \\ \mathbf{0} & \mathbf{V}_{f} \end{bmatrix},
\end{equation}
$\mathbf{I}$ being the identity matrix.\\
Carrying out a projection and reduction of the mass matrix $\mathbf{M}$ starting with $\mathbf{V}_{sol}$ leads to
\begin{equation}
    \mathbf{M}_{r} = \mathbf{V}_{sol}^{T} \mathbf{M} \mathbf{V}_{sol} = \begin{bmatrix} \mathbf{V}_{s}^{T} \mathbf{M}_{s} \mathbf{V}_{s} & \mathbf{V}_{s}^{T} \mathbf{M}_{s} \mathbf{P} \\ (\mathbf{P}^{T} \mathbf{M}_{s} + \mathbf{C}_{sf}) \mathbf{V}_{s}  & \mathbf{M}_{f} + (\mathbf{P}^{T} \mathbf{M}_{s} + \mathbf{C}_{sf})\mathbf{P} \end{bmatrix}.
\end{equation}
Simplifying the matrix as follows
\begin{equation}
    \mathbf{M}_{rs} = \mathbf{V}_{s}^{T} \mathbf{M}_{s} \mathbf{V}_{s},\qquad \mathbf{H}_s = \mathbf{P}^{T} \mathbf{M}_{s} + \mathbf{C}_{sf},\qquad \mathbf{M}_{f,2} = \mathbf{M}_{f} + \mathbf{H}_s \mathbf{P}
\end{equation}
yields the reduced mass matrix
\begin{equation}
    \mathbf{M}_{r} = \begin{bmatrix} \mathbf{M}_{rs} & \mathbf{V}_{s}^{T} \mathbf{M}_{s} \mathbf{P} \\ \mathbf{H}_{s} \mathbf{V}_{s}  & \mathbf{M}_{f,2} \end{bmatrix}.
\end{equation}
With the reduced matrix in this form, it becomes clear that the reduction has only been applied to the structural domain so far. In the weakly coupled bases approach the not yet reduced $\mathbf{M}_{f,2}$ would typically be reduced to $\mathbf{M}_{rf}$ using the projection matrix $\mathbf{V}_{f}$ resulting in
\begin{equation}
    \mathbf{M}_{rf} = \mathbf{V}_{f}^{T} \mathbf{M}_{f,2} \mathbf{V}_{f}.
\label{eq:strongly_coupled_approach_5}
\end{equation}
However, it becomes evident that the projection matrix $\mathbf{V}_{f}$ should be based on the updated mass matrix $\mathbf{M}_{f,2}$, rather than the original mass matrix $\mathbf{M}_{f}$, to account for the apparent additional mass caused by the coupling. Thus, the projection matrix for the fluid domain $\mathbf{V}_{f,2}$ is determined by solving the eigenvalue problem of the updated mass matrix for a subset of $k_{f}$ eigenvectors
\begin{equation}
    (\mathbf{K}_{f} + s^2 \mathbf{M}_{f,2}) \mathbf{p} = \mathbf{0}.
\end{equation}
The sought projection matrix can therefore be expressed as
\begin{equation}
    \mathbf{V} = \begin{bmatrix} \mathbf{V}_s & \mathbf{P}\mathbf{V}_{f,2} \\ \mathbf{0} & \mathbf{V}_{f,2} \end{bmatrix}.
\end{equation}
In contrast to the weakly coupled bases approach, this method accounts for the coupling effects in both the structural and fluid domains.

\begin{figure}[H]
	\centering
	\includegraphics[width=0.9\linewidth]{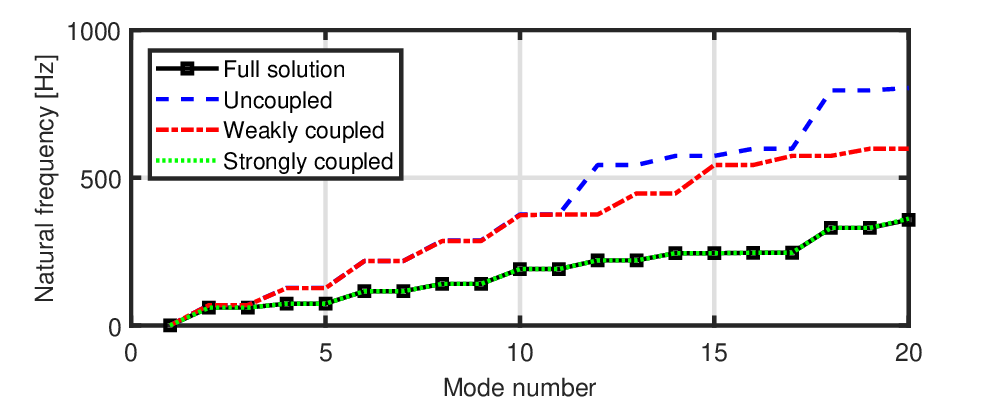}
	\caption{Comparison of the eigenfrequencies of the first 20 modes using different modal methods. Only the strongly coupled basis leads to matching frequencies for all modes, the other bases skip several modes predicted by the full solution in this case of heavy fluid loading.}
    \label{Fig:ucb_wcb_scb_comp}
\end{figure}

A first comparison of these three modal based methods, using the benchmark model provided in Section~\ref{sec:water-filled_cylinder}, is given in Fig.~\ref{Fig:ucb_wcb_scb_comp}. The natural frequencies of the original FSI system are approximated by applying the different approaches. As is evident, only the strongly coupled bases method (SCB) is capable of matching all the natural frequencies beyond the first three modes, implying that correctly addressing the coupling between the two domains in the reduction process is crucial. The uncoupled bases and weakly coupled bases can therefore be considered to be unsuitable for a coupled dynamic system involving a heavy fluid. However, they can still serve as a starting point for other more sophisticated methods, as we will see in the next paragraphs.

\paragraph{Uncoupled bases extended by residual modes iteration (RMI)}
This method is an extension to the uncoupled bases approach. Starting point is again the calculation of a subset of eigenvectors in both the structural and fluid domains, leading to a projection matrix $\mathbf{V}$ as outlined in \Eq{eq:projection_matrix}. The projection matrix is then used to reduce the system matrices $\mathbf{M}$ and $\mathbf{K}$ to $\mathbf{M}_{r}$ and $\mathbf{K}_{r}$.\\
\\
The homogeneous eigenvalue problem in the reduced space is written as
\begin{equation}
    (\mathbf{K}_{r} + s^2 \mathbf{M}_{r}) \begin{bmatrix} \mathbf{u}_{r} \\ \mathbf{p}_{r}\end{bmatrix} = \mathbf{0}.
\end{equation}
Using the projection matrix $\mathbf{V}$, an approximation of the exact modal solution can be obtained by
\begin{equation}
    \begin{bmatrix} \mathbf{u} \\ \mathbf{p}\end{bmatrix} \approx \begin{bmatrix} \hat{\mathbf{u}} \\ \hat{\mathbf{p}}\end{bmatrix} = \mathbf{V} \begin{bmatrix} \mathbf{u}_{r} \\ \mathbf{p}_{r}\end{bmatrix} = \begin{bmatrix} \mathbf{V}_s & \mathbf{0} \\ \mathbf{0} & \mathbf{V}_{f} \end{bmatrix} \begin{bmatrix} \mathbf{u}_{r} \\ \mathbf{p}_{r}\end{bmatrix}.
\end{equation}
The approximated eigenvectors can be used to calculate a resulting force, known as the force residual $\mathbf{R}_{F}^{i}$, for each eigenfrequency $s_i=\omega_{i}$
\begin{equation}
    \mathbf{R}_{F}^{i} = (\mathbf{K} - s_{i}^{2} \mathbf{M}) \begin{bmatrix} \hat{\mathbf{u}}_{i} \\ \hat{\mathbf{p}}_{i}\end{bmatrix}.
\end{equation}
The force residuals can be separated into contributions from the structural domain and the fluid domain and converted into displacement and pressure residues
\begin{equation}
\label{eq:force_residual_expensive}
\begin{split}
    \mathbf{R}_{s}^{i} &= \mathbf{K}_{s}^{-1} \mathbf{R}_{F,s}^{i} \\ 
    \mathbf{R}_{f}^{i} &= \mathbf{K}_{f}^{-1} \mathbf{R}_{F,f}^{i}.
\end{split}
\end{equation}
The obtained displacement and pressure residuals $\mathbf{R}_{s,f} = \biggl[\mathbf{R}_{s,f}^{1}, \mathbf{R}_{s,f}^{2},...,\mathbf{R}_{s,f}^{k_{s},k_{f}}\biggr]$ are now added to the projection matrix $\mathbf{V}$, improving its accuracy
\begin{equation}
    \mathbf{V} = \begin{bmatrix} \mathbf{V}_{s} & \mathbf{0} & \mathbf{R}_{s} & \mathbf{0} \\ \mathbf{0} & \mathbf{V}_{f} & \mathbf{0} & \mathbf{R}_{f} \end{bmatrix}.
\end{equation}
With the extended projection matrix, a new set of residuals can be calculated, further improving the model reduction. This process is repeated until an error criterion is met and additional iterations do not lead to further improvements.\\

\paragraph{Iterative reduced correction algorithm (IRCA)}
The IRCA algorithm, introduced by~\citet{Tabak}, is an improved residual-based method that incorporates correction vectors. In contrast to the RMI approach, the IRCA method additionally enriches the basis $\mathbf{V}$ with correction vectors $\mathbf{X}$ that directly capture the coupling effects between the structure and fluid domains. This leads to fewer iterations and a reduced computational cost.\\
To calculate a correction vector $\mathbf{x}_{s}^{i}$ in the structural domain, the eigenvector $\mathbf{v}_{f,i}$ is assumed to be the correct solution for the fluid domain. With this component fixed, one can now determine the accurate solution in the structural domain, which corresponds to the sought correction vector $\mathbf{X}_{s}^{i}$. This process includes coupling effects and is summarized as 
\begin{equation}
    \Bigg( s^2 \begin{bmatrix} \mathbf{M}_s & \mathbf{0} \\ \mathbf{C}_{sf} & \mathbf{M}_f \end{bmatrix}
    +\begin{bmatrix} \mathbf{K}_s & \mathbf{C}_{fs} \\ \mathbf{0} & \mathbf{K}_f \end{bmatrix}
    \Bigg)
    \begin{bmatrix} 
    \mathbf{x}_{s}^{i} \\ \mathbf{v}_{f,i}
    \end{bmatrix} 
    = 
    \begin{bmatrix} 
    \mathbf{0} \\ \mathbf{0}
    \end{bmatrix}.
\end{equation}
For each eigenvector $\mathbf{v}_{f,i}$ in the fluid domain, a corresponding correction vector $\mathbf{X}_{s}^{i}$ in the structural domain is computed. As known, the projection matrix $\mathbf{V}_{f}$ holds all eigenvectors $\mathbf{v}_{f,i}$ as columns. The correction matrix $\mathbf{X}_{s}$ similarly contains all structural domain correction vectors. It follows
\begin{equation}
\label{eq:IRCA_cost}
    \mathbf{X}_{s} = -\biggl(\mathbf{K}_{s} + s^2 \mathbf{M}_{s}\biggr)^{-1} \mathbf{C}_{fs} \mathbf{V}_{f}.
\end{equation}
The correction vectors in the fluid domain $\mathbf{X}_{f}$ can be determined accordingly by assuming the eigenvector $\mathbf{v}_{s,i}$ as the correct solution for the structural domain.\\
With the residual vectors calculated in the same manner as described for the residual modes iteration approach, the projection matrix is then constructed as:
\begin{equation}
    \mathbf{V} = \begin{bmatrix} \mathbf{V}_{s} & \mathbf{0} & \mathbf{R}_{s} & \mathbf{0} & \mathbf{X}_{s} & \mathbf{0} \\ \mathbf{0} & \mathbf{V}_{f} & \mathbf{0} & \mathbf{R}_{f} & \mathbf{0} & \mathbf{X}_{f} \end{bmatrix}.
\end{equation}
After orthogonalization of the basis $\mathbf{V}$, the residual and correction vectors are recalculated iteratively until a certain error criterion is met.
The symmetric formulation of the problem to enhance the computational efficiency presented previously cannot be straightforwardly applied to the IRCA procedure or modal method based MOR techniques in general. To retain the computational advantages offered by a symmetrized problem, Tabak et al. presented in \cite{Tabak_sym} a symmetrization process for modal methods such as the IRCA procedure.\\
The symmetrization process for the IRCA method is briefly outlined.
By using the transformation matrix $\mathbf{\tau}$
\begin{equation}
    \mathbf{\tau} = \begin{bmatrix} \mathbf{K}_{s}^{-1}\mathbf{M}_{s} & -\mathbf{K}_{s}^{-1}\mathbf{C}_{fs} \\ \mathbf{0} & \mathbf{I} \end{bmatrix},
\end{equation}
a change of variables is performed
\begin{equation}
    \begin{bmatrix} \mathbf{u} \\ \mathbf{p} \end{bmatrix} = \mathbf{\tau} \begin{bmatrix} \mathbf{\hat{u}} \\ \mathbf{p} \end{bmatrix}.
\end{equation}
The change of variables will result in the following system
\begin{equation}
    \begin{bmatrix} \mathbf{M}_s\mathbf{K}_{s}^{-1}\mathbf{M}_{s} & -\mathbf{M}_s\mathbf{K}_{s}^{-1}\mathbf{C}_{fs} \\ \mathbf{C}_{sf}\mathbf{K}_s^{-1}\mathbf{M}_s & (\mathbf{M}_f-\mathbf{C}_{sf}\mathbf{K}_s^{-1}\mathbf{C}_{fs}) \end{bmatrix} \begin{bmatrix} \mathbf{\hat{\ddot u}} \\ \mathbf{\ddot p} \end{bmatrix} + \begin{bmatrix} \mathbf{M}_{s} & \mathbf{0} \\ \mathbf{0} & \mathbf{K}_f \end{bmatrix} \begin{bmatrix} \mathbf{\hat{u}} \\ \mathbf{p} \end{bmatrix} = \begin{bmatrix} \mathbf{f}_{s} \\ \mathbf{f}_f \end{bmatrix}.
\end{equation}
Since $(-\mathbf{M}_s\mathbf{K}_{s}^{-1}\mathbf{C}_{fs})^T = \mathbf{C}_{sf}(\mathbf{K}_s^{-1})^{T}\mathbf{M_{s}^{T}}$ holds, it is evident that the system matrices are now symmetric.\\

\subsubsection{Krylov subspace method}
Unlike modal reduction, which constructs an invariant subspace from selected (generalized) eigenvectors and is therefore tied to the system's eigenvalues, Krylov-based methods generate the basis by repeated multiplication of the system matrix with an input vector, forming a polynomial subspace.
To target specific frequency ranges, rational Krylov subspaces exploit shift-invariance, enabling the construction of bases centered around chosen expansion points. This property allows the reduced model to accurately capture the input--output behavior within a frequency band of interest~\cite{Saad_2003}.
Krylov subspaces achieve this by matching moments of the transfer function at these frequencies, which makes them particularly suitable for forced-response and parametric problems. In contrast, modal methods approximate the entire system dynamics, including components unrelated to the transfer path under investigation, leading to unnecessary computational effort for large-scale systems with a known excitation point. Krylov subspaces are directly applicable to linear first order systems. Second order system can be reformulated into an equivalent first-order state-space representation introducing displacement and velocity as state variables leading to the second order Krylov sequence. This approach was initially introduced by \citet{Su}, and later refined \cite{Bai2,Bai1,Bai3}.
The procedure is outlined as follows:\\
The solution to the second order system, which reads as
\begin{equation}
      \mathbf{x}(s) = (s^2\mathbf{M} + s\mathbf{D}+\mathbf{K})^{-1}\mathbf{f}
\end{equation}
can be expanded around the point $s_0$ as follows
\begin{equation}
       \mathbf{x}(s) = \biggl((s-s_0)^2\mathbf{M} + (s-s_0)\mathbf{D}+\mathbf{K} +2ss_0\mathbf{M}+s_0\mathbf{D}-s_0^2\mathbf{M}\biggr)^{-1}\mathbf{f} .
\end{equation}
By summarizing the terms as
\begin{equation}
    \tilde{\mathbf{D}} = \mathbf{D} + 2 s_0 \mathbf{M}
\end{equation}
\begin{equation}
    \tilde{\mathbf{K}} = \mathbf{K} + s_0\mathbf{D} + s_0^2\mathbf{M},
\end{equation}
the displacement can be rewritten as
\begin{equation}
        \mathbf{x}(s) = \biggl((s-s_0)^2\mathbf{M} + (s-s_0)\tilde{\mathbf{D}}+\tilde{\mathbf{K}}\biggr)^{-1}\mathbf{f}.
\end{equation}
The transfer function around the expansion point $s_0$ can be written as a series expansion and its leading moments can be approximated by projecting the system onto the following second order Krylov subspace 
\begin{equation}
        \mathscr{K}_{k}
(\mathbf{P},\mathbf{Q},\mathbf{r}_0) := \text{span}(\mathbf{r}_0,\mathbf{r}_1,...,\mathbf{r}_{k-1})
\end{equation}
with
\begin{equation}
\label{eq:Krylov_expenses}
\begin{split}
        \mathbf{r}_0 &= \tilde{\mathbf{K}}^{-1}\mathbf{f} \\
        \mathbf{r}_1 &= \mathbf{P}\mathbf{r}_0 \\
        \mathbf{r}_k &= \mathbf{P}\mathbf{r}_{k-1} + \mathbf{Q}\mathbf{r}_{k-2},
\end{split}
\end{equation}
where $\mathbf{P} = -\tilde{\mathbf{K}}^{-1} \tilde{\mathbf{D}}$ and $\mathbf{Q} = -\tilde{\mathbf{K}}^{-1}\mathbf{M}$. This Krylov subspace is referred to as the second order Krylov subspace of order $k$, and the sequence $\{\mathbf{r}_k\}$ used to span it is known as the second order Krylov vector sequence of order $k$. The projection basis is derived from the second order rational Krylov subspace using the Arnoldi algorithm, resulting in an orthonormal projection matrix $\mathbf{V} \in \mathbb{R}^{n \times k}$.

\subsection{Computational complexity due to the matrix inversion}
In the following section, the effectiveness of the four most accurate MOR methods will be compared in detail. It can already be observed that the SCB method requires the explicit calculation of $\mathbf{P} = -\mathbf{K}_s^{-1}\mathbf{C}_{fs}$ in  \Eq{eq:schur_complement_2}, solving a system of equations with respect to $\mathbf{C_{fs}}$ or doing a matrix decomposition for $\mathbf{PV}_{f,2}$, see \Eq{eq:PV_f_decomposition}, which both are computationally expensive. Although the RMI approach operates on a relatively small residual matrix (e.g., $\mathbf{R}_{F,s}^{i}$), the computation of residuals in \Eq{eq:force_residual_expensive}
$\mathbf{R}_{s}^{i} = \mathbf{K}_{s}^{-1}\mathbf{R}_{F,s}^{i}$ and their iterative updates significantly increase the overall computational complexity. The same holds for the IRCA method, where the additional calculation of the correction vectors significantly raises the computational cost, see \Eq{eq:IRCA_cost}. Finally, the computation of the TOAR algorithm  requires only a Cholesky decomposition of $\mathbf{r_0, P, Q}$ once, see \Eq{eq:Krylov_expenses} and can be reused for the basis construction. Since these three variables are frequency independent, these enable significant advantages to a full FE solver, requiring matrix factorization at each frequency step. Based on these initial insights, this work aims at finding the limits of the proposed methods, pointing out why for large systems the Krylov approach is the only one resulting in a computationally feasible solution.

\section{Benchmark case: water-filled cylinder}
\label{sec:water-filled_cylinder}
As a benchmark model, we introduce a water-filled Plexiglas cylinder. This simple geometry allows for controlled mesh generation using hexahedral elements, while remaining sufficiently complex due to the asymmetry between applied forces, internal pressure, and acceleration. Additionally, the boundary conditions are well defined. 

\subsection{Experimental setup}
\begin{figure}[h!]
    \centering
	\includegraphics[width=1\linewidth]{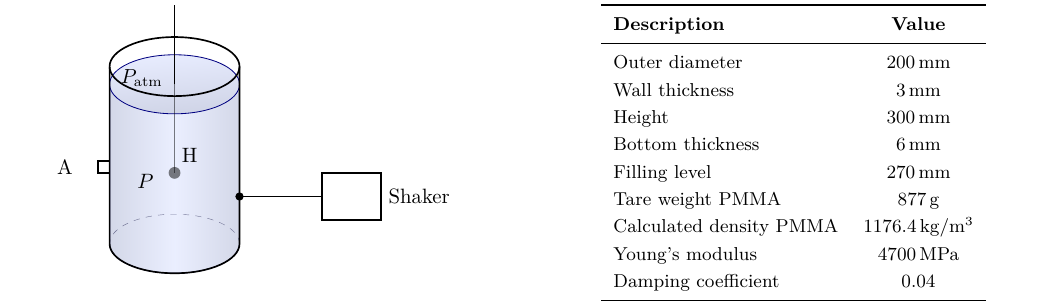}

    \caption{Left: Experimental setup of the water-filled cylinder including measurement positions of the accelerometer (A) and the hydrophone (H). Right: List of material properties of the cylinder.}
    \label{fig:experimental_setup_combined}
\end{figure}

Experiments were carried out using a water-filled elastic cylinder to examine the FSI problem in a controlled and simplified environment. The resulting experimental database includes frequency response functions, which serve as a reference for model validation. The scheme of the experimental setup is given in Fig.~\ref{fig:experimental_setup_combined}. A shaker is used to excite the cylinder. The sensor (A) measures the accelerations in the structural domain and a hydrophone (H) sensor measures the pressure (P) in the fluid domain.
The hollow cylinder with a closed bottom was filled up to a certain level with water to simulate a water-filled enclosure. The cylinder is made of Plexiglas (PMMA) and its relevant dimensions are illustrated in the table of Fig.~\ref{fig:experimental_setup_combined}.

\begin{figure}[h!]
  \centering
	\includegraphics[width=1\linewidth]{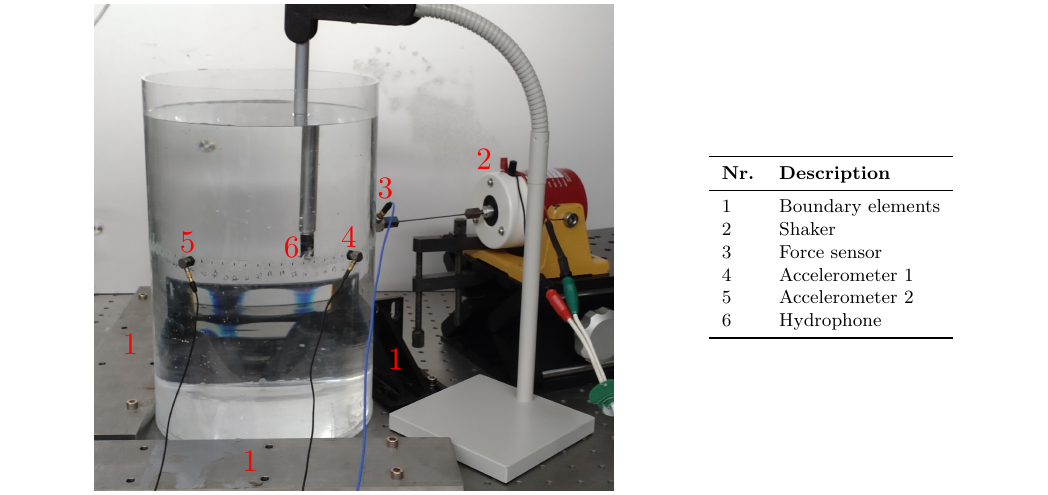}
  \caption{The experimental setup for the water-filled cylinder on the vibration isolation table,  illustrating boundary conditions, measurements devices and the type and position of the excitation. }
   \label{Fig:Experimental_setup}
\end{figure}

The experimental setup is shown in Fig.~\ref{Fig:Experimental_setup}. The cylinder was placed on a vibration isolation table and clamped in position using boundary elements (1) to prevent any radial displacements at the bottom. A shaker (2) was used to excite the cylinder. A force sensor (3) was inserted between the anchor point of the cylinder and the stinger of the shaker to characterize the force input into the system. To measure the vibrations of the cylinder caused by the excitation, two accelerometers (4, 5) were mounted on the surface of the cylinder. Water was filled to a height of 270\,mm into the cylinder. A hydrophone (6), clamped to a bracket, was used to measure the pressure level within the water at a specific position.
The force was applied at a height of 160\,mm above ground level, acting perpendicular to the excitation surface. Relative to this input position, the first accelerometer (4) was installed at the same height and at a circumferential distance of 110\,mm, and the second accelerometer (5) at a circumferential distance of 260\,mm. The hydrophone (6) was positioned near the first accelerometer (4), in 1\,cm radial distance to the cylinder wall, at the same height as the accelerometer (4). 
Logarithmic sine sweeps, ranging from 10\,-\,10\,000\,Hz, were applied as voltage inputs to the shaker. The force sweep measured by the force sensor was then used as the system input. With the acceleration measurements on the structure and the pressure level measurement in the fluid, serving as system outputs, three frequency transfer functions were measured.

\subsection{Computational setup}

\begin{figure}[ht]
    \centering
    \includegraphics[width=1\linewidth]{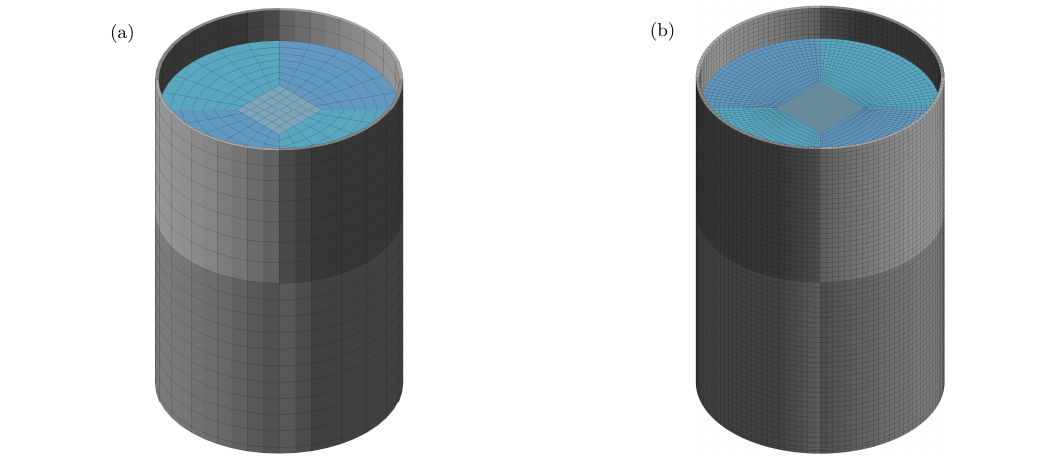}

    \caption{Illustration of two FE models that are investigated in this work. Shown are the 30k (a) and 1000k (b) DOF systems, including both the fluid and solid domain.}
    \label{fig:DOF_system_comparison}
\end{figure}

The geometry and mesh of the cylinder were developed in \textsc{Ansys}. The PMMA cylinder was assigned material properties including a Young’s modulus of \(4.7\,\text{GPa}\), a Poisson’s ratio of 0.375, and a density of \(1177\,\text{kg/m}^3\). The water inside the cylinder was modeled with a density of \(998.2\,\text{kg/m}^3\) and a speed of sound of \(1482.1\,\text{m/s}\). Radial constraints with zero displacement at the bottom of the cylinder's lateral area  were applied to represent the boundary conditions imposed  to the boundary elements in the experimental setup. The pressure at the open surface of the fluid was set to zero.  The force was applied on a single node, normal to the surface direction at the same location as the shaker. In order to investigate the scalability and efficiency of the MOR methods, five different mesh sizes were generated in Ansys 2024 R2, resulting in five different coupled FSI systems with a total number of DOFs of approximately 10k, 30k, 100k, 300k, and 1000k. These five systems were used as benchmark for the comparative study of the different MOR techniques. Fig.~\ref{fig:DOF_system_comparison}a and~\ref{fig:DOF_system_comparison}b illustrate the mesh refinement of the 30k and 1000k DOF systems, respectively. \\
The system matrices $\mathbf{M}$ and $\mathbf{K}$ were generated for each system. These system matrices, the respective nodes, and their coordinates were then exported to Matlab for further calculations, including the computation of full FE solutions and MOR. All calculations in Matlab were performed on a system equipped with an Intel® Xeon® Gold 6248R processor (24 cores, 3.00\,GHz), 768\,GB RAM, and a 64-bit operating system.

\section{Results}
\label{sec:results}
\subsection{Comparison of experimental and full model}
\begin{figure}[h!]
  \centering
 \includegraphics[width=1\linewidth]{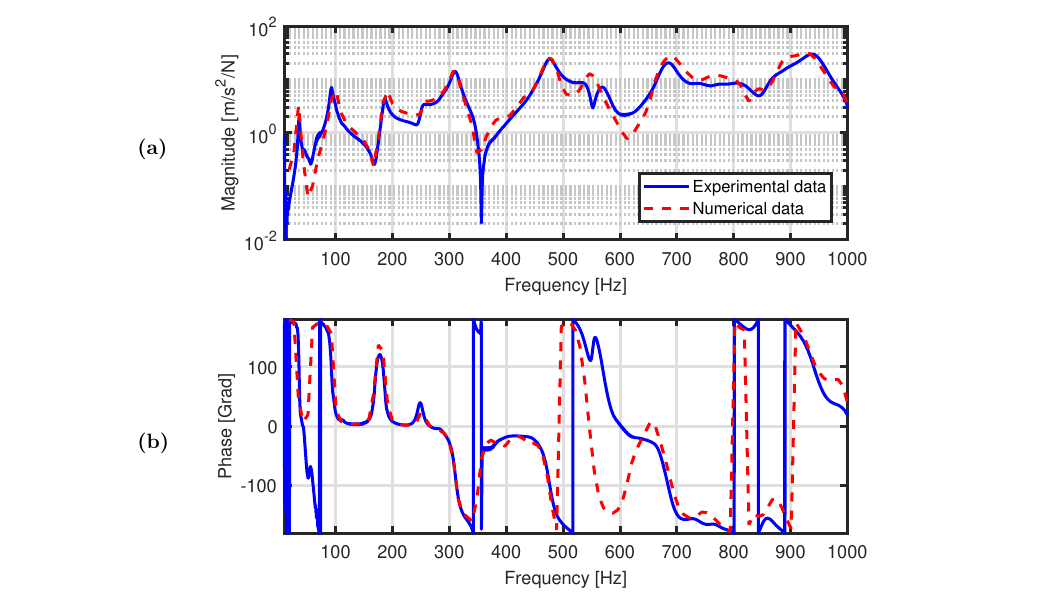}

  \caption{Structural domain validation: The magnitude of the frequency response function (a) and its phase (b) are shown for the transfer path between the input force and the accelerometer at position 4 in Fig.~\ref{Fig:Experimental_setup}.}
  \label{fig:Validation_acc1_10-1000Hz}
\end{figure}

A comparative study was conducted for all different system sizes of the cylinder model. Convergence is already achieved with the 30k DOF system for frequencies below 1000\,Hz. For the experimental validation of the model, the numerical results of the 1000k DOF system were used.
Fig.~\ref{fig:Validation_acc1_10-1000Hz} shows the experimentally measured and the numerically calculated transfer function between the input force and the output acceleration of accelerometer 1 in the frequency range of 1–1000\,Hz. The calculated data correlate well with the experimental data up to 1000\,Hz. The largest discrepancy occurs around 550\,Hz, where a resonance of the numerical model corresponds to an antiresonance in the measured system due to proximity to a nodal line. The phase of the acceleration equally demonstrates good agreement.\\

\begin{figure}[h!]
  \centering
  \includegraphics[width=1\linewidth]{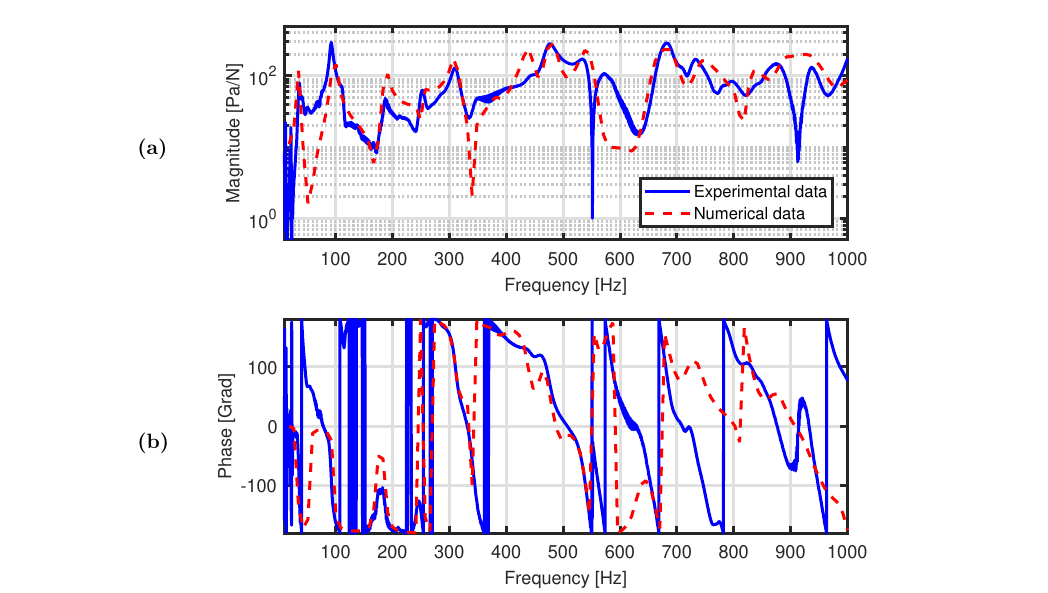}

  \caption{Fluid domain validation: The magnitude of the frequency response function of the pressure (a) and its phase (b) are shown for the transfer path between the input force and the hydrophone at position 6 in Fig.~\ref{Fig:Experimental_setup}.}
  \label{fig:Validation_p4_10-1000Hz}
\end{figure}
Fig.~\ref{fig:Validation_p4_10-1000Hz} shows the magnitude and phase of the transfer function between the input force and the output pressure. The data show a reasonable correlation within the frequency range of 150 to 500\,Hz. At higher frequencies, however, the agreement between experimental and numerical results weakens. Nevertheless, the magnitudes of the experimental and numerical data remain within the same order of magnitude throughout the frequency range.
With these results, the numerical models can be regarded as sufficiently validated for the purpose of using them for a comparative MOR study. To further improve the fit between the experimental and numerical data, the experiment would need to be conducted under more ideal boundary conditions, which is beyond the scope of this work. Given the mass of the system, the clamps can no longer be considered rigid at higher frequencies, which considerably complicates the model. 
\subsection{Visualization of mode shapes}

\begin{figure}[h!]
    \centering
    \includegraphics[width=1\linewidth]{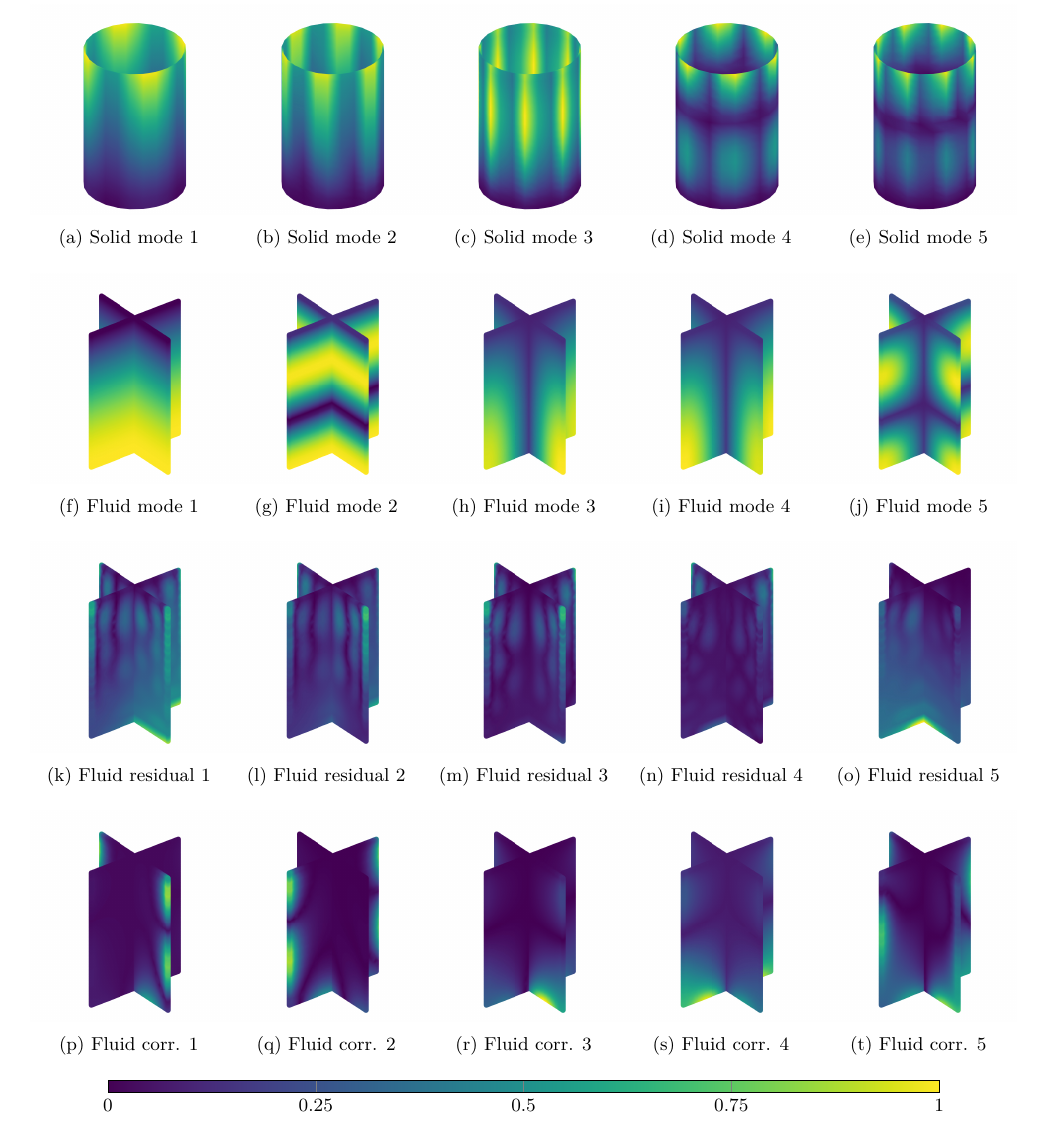}
    
    \caption{Visualization of the first five normalized solid eigenmodes (a-e), fluid eigenmodes(f-j), residuals (k-o) used in the RMI method and the correction vectors (p-t) used in the IRCA method, defined both in the fluid domain.}
    \label{Fig:modeshapes}
\end{figure}

To complement the detailed overview of the existing modal methods and coupling techniques in section~\ref{sec:modeling}, we present a comprehensive overview of the solid and fluid modes of the cylinder structure. In the first row of Fig.~\ref{Fig:modeshapes}, the first five eigenvectors of the uncoupled structural system are presented. In the second row, the resonances of the fluid are shown in two perpendicular cross sections. In the third row, the first five residual vectors that were used for the RMI and the IRCA methods are illustrated. Lastly, the correction vectors obtained from the IRCA method are shown. While the modes in the first two rows represent the physical system dynamics at resonance, the last two rows show that the coupling terms are non‑negligible and strongly influence the interaction between the fluid and structural modes. 

\subsection{Comparison of different MOR methods}

\begin{figure}[h!]
	\centering
    \includegraphics[width=1.0\linewidth,trim=10 0 10 0, clip]{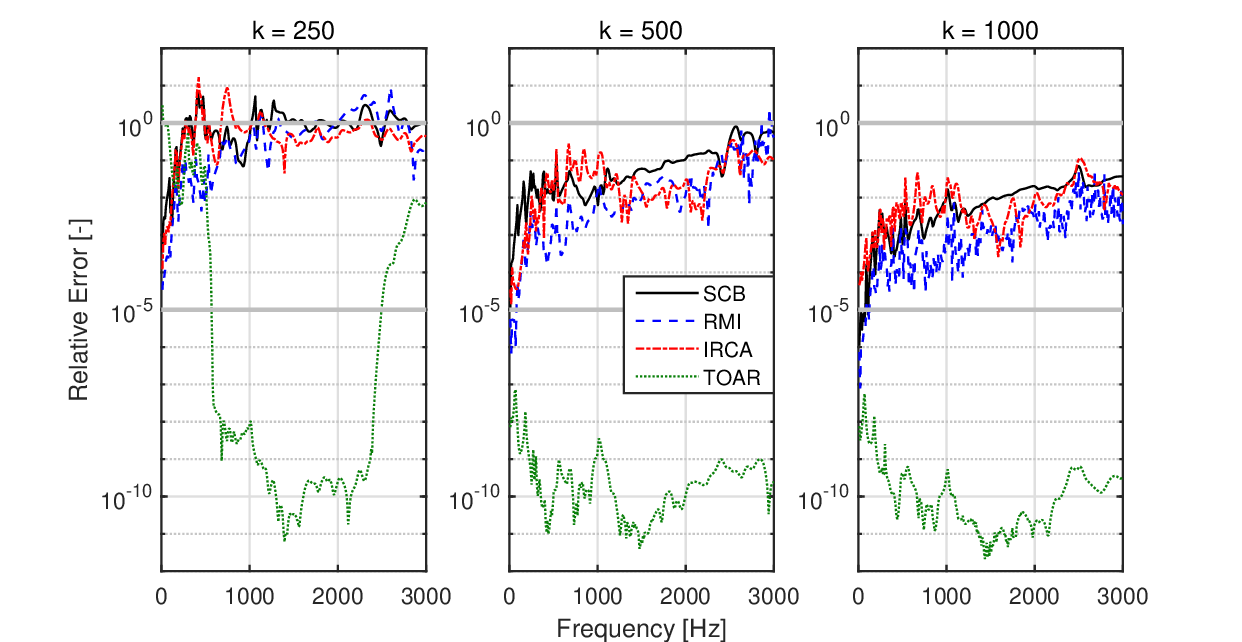}
	\caption{Relative error achieved by different MOR techniques for an increasing number of modes k (left to right).}
    \label{fig:comparison_of_methods}
\end{figure}

The relative errors achieved by the strongly coupled bases approach, the residual modes iteration, the IRCA and the TOAR method, when approximating the same transfer function in the structural domain are illustrated in Fig.~\ref{fig:comparison_of_methods}, to highlight the superior accuracy of the TOAR method. The figure clearly indicates that the TOAR method maintains a relative error of approximately $10^{-10}$ around the chosen expansion point ($f_0 = 1500$\,Hz). This low relative error can be achieved across the entire frequency range for a number of modes $k = 350$. Further increasing $k$ does not significantly reduce the relative error, as illustrated in the figure.

\renewcommand{\arraystretch}{1.2} 

\begin{figure}[h!]
  \centering
      \includegraphics[width=0.88\textwidth]{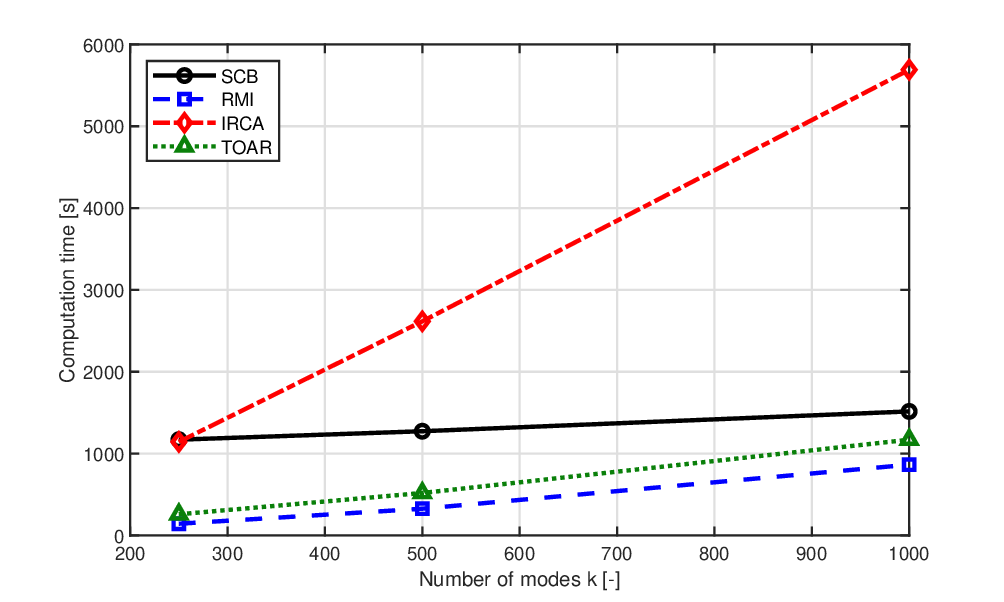}
  \caption{\label{Fig:Comp_times_comparison}Comparison of k-dependent basis computation time for the 30k DOF system for the four investigated MOR schemes.}
\end{figure}

The modal-based MOR techniques do not achieve this range of low relative errors. For a relatively low number of modes $k=250$, the relative error remains around $10^{-1}$. A further increase in $k$ up to 1000 results in relative errors around $10^{-3}$, and for lower frequencies even below $10^{-5}$.\\ However, the approximation precision of the Krylov-based TOAR method is never achieved by these modal methods.
Fig.~\ref{Fig:Comp_times_comparison} illustrates computation times needed for the four different MOR methods to construct a basis with a given number of modes $k$, required to achieve the relative errors shown in Fig.~\ref{fig:comparison_of_methods}. 
It reveals that the TOAR method, together with the residual modes iteration method, is computationally the most efficient. The basis computation time increases linearly with the number of modes $k$. For the TOAR method, this behavior can be explained by the second order Krylov sequence. Specifically, each increment in $k$ results in the computation of an additional second order Krylov vector, which is derived from the two previously computed vectors to enrich the basis. Therefore, the computational cost increases in a steady and predictable manner.\\
The same holds for the residual modes iteration, where for an increasing number of modes $k$, additional residuals need to be computed, again leading to a linear and predictable increase in computational cost.
The strongly coupled bases method has a high initial computational cost due to factorizations of the stiffness matrices in both the structural and fluid domains, $\mathbf{K}_s$ and $\mathbf{K}_f$. Once these operations are completed, the remaining computations require minimal time. Therefore, increasing the value of $k$ is manageable without significant additional computational cost.
The IRCA method exhibits the highest computational cost because the correction vectors must be computed explicitly. These vectors need to be computed $k$ times in each iteration, leading to a steep increase in computation time.\\

Tab.~\ref{Tab:Feasibility} shows the results of the feasibility study. As evident from the table, the Krylov subspace based TOAR technique and the modal methods based uncoupled bases approach were the only methods capable of computing a projection basis even for the largest systems, with up to 1000k DOF. In contrast, all other techniques reached their limit at the system of 100k DOF, due to limited computing machine capabilities and exponentially growing computational costs. Since only the Krylov approach provides high accuracy for the harmonic forced vibration problem over the entire frequency range, it is the recommended method. However, the drawback is that the basis is only valid for the selected excitation position and direction. Different load cases require different bases, which may lead to unacceptable computational costs.
\begin{table}[h!]
    \centering
    \small
    \renewcommand{\arraystretch}{1.2} 
    \setlength{\tabcolsep}{8pt}       
    \caption{Feasibility of different methods for FSI models of varying sizes of DOF.}
    \begin{tabular}{c c c c c c}
        \toprule
        \diagbox[width=3.2cm, height=1.0cm]{\textbf{DOF}}{\textbf{Method}}  
        & \textbf{SCB} & \textbf{RMI} & \textbf{IRCA} & \textbf{TOAR} & \textbf{Ratio solid/total DOF } \\
        \midrule
        10k (9,945) & \ding{51} & \ding{51} & \ding{51} & \ding{51} & 0.776 \\
        30k (28,000) & \ding{51} & \ding{51} & \ding{51} & \ding{51} & 0.512 \\
        100k (105,614) & \ding{51} & \ding{51} & \ding{51} & \ding{51} & 0.437\\
        300k (293,394) & \ding{55} & \ding{55} & \ding{55} & \ding{51} & 0.346\\
        1000k (954,504
) & \ding{55} & \ding{55} & \ding{55} & \ding{51} & 0.284 \\
        \bottomrule
    \end{tabular}
    \label{Tab:Feasibility}
\end{table}

Finally, we outline the achieved speed-up of the calculations by using the TOAR method compared to high fidelity FE computations. Tab.~\ref{tab:mesh_time_study} reports timing results for frequency‑domain analyses of the five mesh sizes. ANSYS provides the reference full FE solution with an average per‑frequency time. In MATLAB the computational costs of the TOAR algorithm (including initial costs of Cholesky decomposition and basis construction) and the full FE solution was computed per frequency and extrapolated for frequency sweep of 1000 frequency steps. The MATLAB results are separated into the contributions of the TOAR algorithm including the one‑time Cholesky factorization and the TOAR subspace construction for three ROM orders ($k$ = 10, 20, 30) as well as the full FE solution was computed per frequency and extrapolated for frequency sweep including 1000 frequency steps. For a direct comparison within Matlab, the speedup factor is computed and shows for large systems accelerated computations of 560 to 620.  The speedup is defined as the ratio between the time required by a 1000‑point sweep with the full FE model and the time required to build the ROM (factorization plus TOAR construction).
By looking solely at the largest model with 1000k DOF, a single frequency step calculation takes in average 7071 seconds (1.96 hours) for solving the FE system. The construction time of a basis using the TOAR algorithm with 250 modes requires 14.43 hours. This scenario becomes particularly relevant when employing, for instance, ten expansion points with 25 mode vectors each, or when extending the approach to parametric model order reduction (pMOR).
As mentioned before, this entire study has been conducted using Matlab. Commercial FEM solvers are likely faster since they are optimized for the limited amount of different matrix calculations. This proves to be true when looking at the times measured in \textsc{Ansys}. For completeness, we show the full solution calculation time in \textsc{Ansys} which is approximately 5 times faster than Matlab.
The comparison with \textsc{Ansys} shows that the Krylov subspace approach is already more than two times faster for 100 frequency steps.
\begin{table}[h!]

\centering
    \caption{Performance comparison for different mesh sizes and computational costs. The speedup is calculated based on the ratio between the summation of the Cholesky decomposition and the time to compute a basis with $n=30$ basis vectors and the estimated total time to compute 1000 frequency steps explicitly. }

\begin{tabular}{
            |>{\raggedright\arraybackslash}p{1.2cm}| >{\centering\arraybackslash}p{1.5cm}
              | >{\centering\arraybackslash}p{1.1cm}
              >{\centering\arraybackslash}p{1.2cm}
            >{\centering\arraybackslash}p{1.2cm}
              >{\centering\arraybackslash}p{1.2cm}
              >{\centering\arraybackslash}p{1.3cm} 
              >{\centering\arraybackslash}p{1.4cm}
              >{\centering\arraybackslash}p{1.4cm}|
              >{\centering\arraybackslash}p{1.4cm}|
        }
\hline
\textbf{Software} & \textbf{Ansys Full FE} & \multicolumn{7}{c|}{\textbf{MATLAB}} &  \\
\cline{3-8}\cline{9-10}  
\multicolumn{1}{ |>{\raggedright\arraybackslash}p{1.3cm}|}{%
  \diagbox[
    width=1.3\linewidth, 
    innerleftsep=0pt, innerrightsep=0pt
  ]{\color{white}{-}}{\color{white}{-}}%
}                        &       &  \multicolumn{5}{c}{\textbf{TOAR basis calculation time}}      & \multicolumn{2}{|c|}{\textbf{Full FE solution}} &      \\
\textbf{Mesh} & \textbf{per freq. [s]} &  \textbf{Chol. [s]} &  \textbf{k= 10 [s]} &
\textbf{k= 20 [s]} &  \textbf{k= 30  [s]} &  \textbf{k= 250  [s]} & \multicolumn{1}{|c}{ \textbf{ per freq. \newline [s]}} &  \textbf{1,000 freqs [s]} & \textbf{Speedup [-]} \\
\hline
10k  & 1.7   & 3.7   & 2.7    & 5.7    & 8.8 & -  & \multicolumn{1}{|c}{2.3} & 2,260      & 181 \\
30k  & 5.5  & 12.9  & 22.1   & 32.8  & 43.8 &  258  & \multicolumn{1}{|c}{11.9} & 11,900     & 272 \\
100k   & 21.6   & 138.7  & 70.0    & 153.0   & 236.0 & - & \multicolumn{1}{|c}{229.2}  & 229,200    & 610 \\
300k  & 130.7  & 863.4  & 1,213.3  & 1,622.9  & 2,035.0 & - & \multicolumn{1}{|c}{1,259}  & 1,259,900   & 619 \\
1,000k & 1,338 & 5,256 & 2,311.9 & 4,797.4 & 7,357.6 & 46,699 & \multicolumn{1}{|c}{7,071.4} & 7,071,400   & 561 \\
\hline
\end{tabular}
\label{tab:mesh_time_study}

\end{table}

\subsection{Effect of symmetrization and conditioning}
\begin{figure}[h!]
  \centering
  \includegraphics[width=0.88\textwidth]{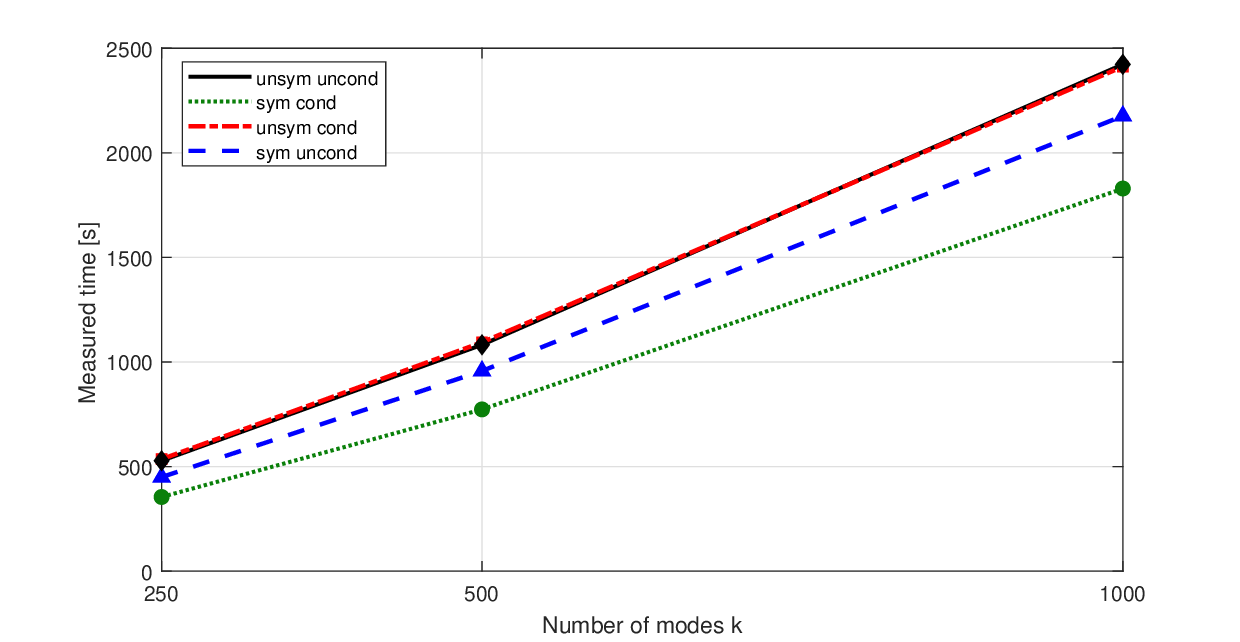}
  \caption{Comparison of $k$-dependent computation times of the TOAR method when investigating conditioning and symmetrization for the 30k‑DOF system.}
 
   \label{Fig:TOAR_sym_cond_time_comparison}
\end{figure}

Symmetrization and conditioning of the system matrices have a significant impact on the performance of the TOAR method in terms of computational efficiency. An overview of is provided in Fig.~\ref{Fig:TOAR_sym_cond_time_comparison}. The figure illustrates the time needed to compute a reduction basis for the 30k DOF FSI system. The TOAR method is applied using $k=250$, $500$, and $1000$, and a single expansion point strategy to investigate the impact.
Non-symmetric matrices exhibit nearly identical computational times, regardless of whether conditioning is applied. In contrast, applying symmetrization leads to a clear reduction in runtime of roughly 10$\%$. The largest gains are achieved when symmetrization is combined with conditioning. This further accelerates the computation, reducing the overall computational time by up to 25$\%$ compared with the non-symmetric unconditioned baseline.

\begin{figure}[h!]
	\centering

    \includegraphics[width=0.99\linewidth,trim=1cm 0cm 1cm 0cm,
    clip]{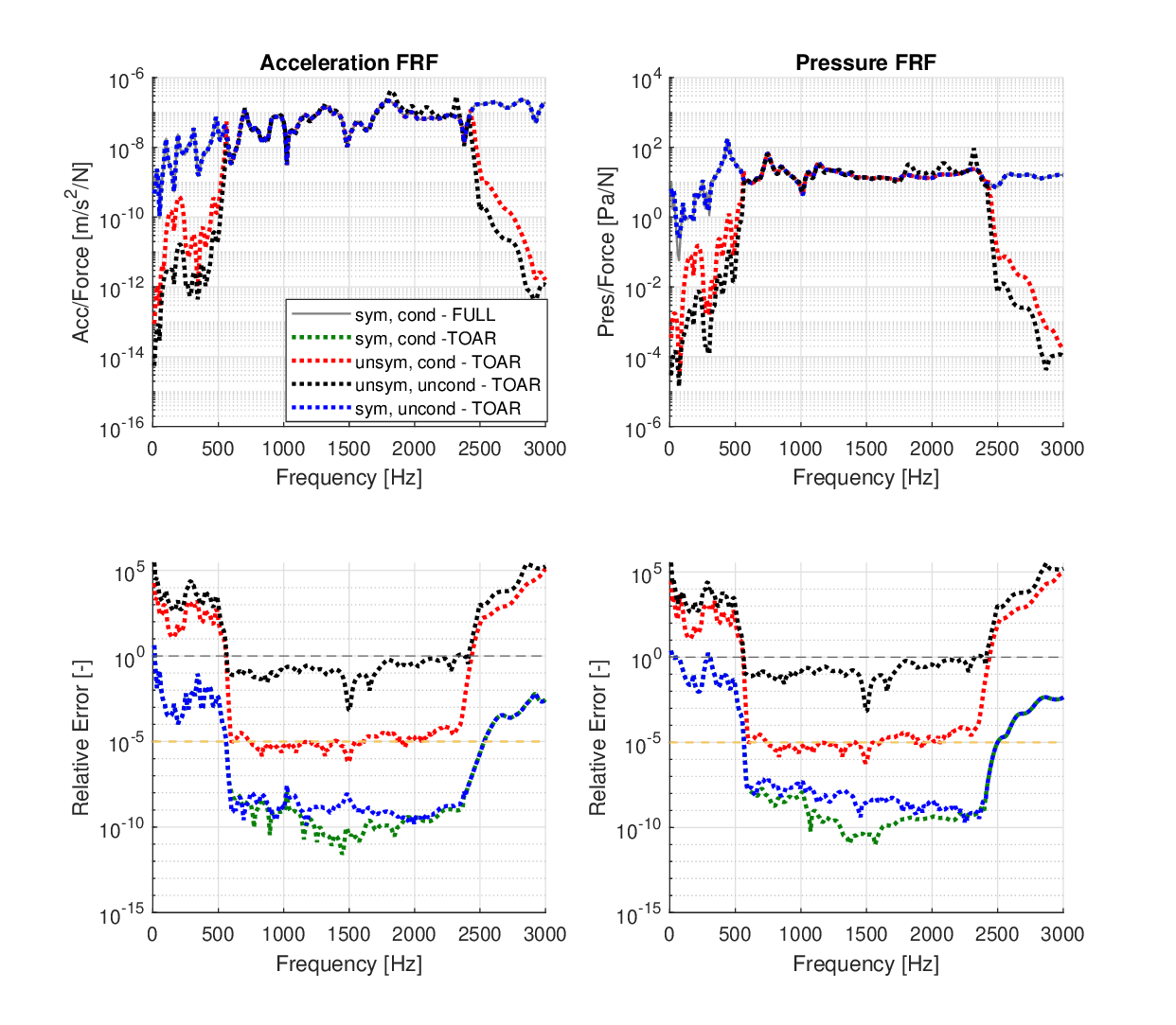}
	\caption{Effect of symmetrization and conditioning when performing the TOAR method with $k=250$ to the 30k DOF system. Shown are the acceleration and pressure FRF approximations for an expansion point at $f_0=1500$ Hz, together with their relative errors.}
    \label{fig:TOAR_sym_cond_overview}
\end{figure}

As illustrated in Fig.~\ref{fig:TOAR_sym_cond_overview}, the symmetrization and conditioning process not only affects the computational time for TOAR but also influences the relative error of the transfer function approximation.\\
If a transfer function between the input force and the radial acceleration at a specific location on the surface of the cylinder is approximated using the TOAR method with non-symmetric and unconditioned system matrices, the relative error remains around $10^{-1}$, and this is only in close proximity to the chosen expansion point at 1500\,Hz.\\
This time, both symmetrization and conditioning individually significantly improve the quality of the approximation and therefore reduce the relative error. Conditioning improves the relative error by a magnitude of $10e-5$. 

However, this time, as opposed to the computational time analysis, once the symmetrization process is applied, an additional conditioning process decreases the relative error only insignificantly.\\
The best result in terms of computational time and accuracy is achieved when both symmetrization and conditioning are combined. This substantially reduces the relative error to around $10^{-11}$, significantly extends the range of a high-quality approximation around the chosen expansion point, and reduces the TOAR basis computation time by a factor of approximately 1.4.

\section{Conclusion}
In this work, we conducted a comprehensive evaluation of reduction strategies for fluid‑coupled vibroacoustic systems. We presented three related modified modal MOR methods and a Krylov subspace projection approach as an alternative reduction scheme. These approaches were then applied as projection-based MOR techniques to a benchmark model from coarse to fine meshes, yielding system sizes from 30k to 1000k DOFs.
The FE models were validated against laboratory experiments up to 1000~Hz. Frequency response functions measured with accelerometers and hydrophones showed good agreement with errors mostly below 5$\%$ compared to the computational predictions across all dominant resonances within the structural and fluid domains. 
The Krylov subspace-based TOAR method demonstrates superior performance in both accuracy and computational efficiency compared to advanced modal-based MOR techniques such as SCB, RMI, and IRCA. While modal approaches require a significantly higher number of modes to achieve comparable accuracy -- often failing to reach the precision of TOAR -- the TOAR method maintains a predictable linear growth in computational cost due to its efficient second-order Krylov sequence. Furthermore, TOAR and RMI exhibit the most favorable scaling behavior with increasing mode count. The feasibility study confirms TOAR as the only method capable of handling systems up to 1000k DOF, making it the most robust and scalable approach. Finally, this method enables substantial speed-ups over full-order FE simulations, achieving acceleration factors exceeding 600$\times$ for large-scale frequency sweeps, thereby establishing it as the recommended technique for efficient and accurate MOR in large-scale FSI problems including heavy fluids. Conditioning and symmetrization of the coupled problem improve the relative error between the ROM and the full FE model by several orders. A combination of both steps, however, does not perform significantly better on the relative error than just applying symmetrization. Nevertheless, for the computational runtime of the TOAR algorithm, a combination of both methods leads to a $25 \%$ reduction of the basis construction time.

Our investigation is limited to selected forced excitation positions; the generated Krylov basis corresponds to that excitation point and may not generalize to arbitrary loading distributions. Extensions to multi-input systems exist, but at significant additional cost~\cite{CHU2008}. Future work will address this by exploring parametric MOR excitation frameworks. More complex geometries with small features result in large-scale dynamical systems with high modal density. Local behavior approximations using Krylov subspaces might still be sufficient to obtain a good approximation of the system~\cite{Gugercin2008,Spescha2018,Brenner2015}.

\section*{Data Availability Statement}
Some or all data, models, or code generated or using during the study are available in a repository online in accordance with the funder data retention policies. The data are available at \href{https://doi.org/10.5281/zenodo.18481410}{https://doi.org/10.5281/zenodo.18481410}~\cite{taenzer_2026_18481410}. 

\section*{Acknowledgments}
This work was supported by grant 213127 of the Swiss National Science Foundation. 

\section*{Competing interests}
The authors have no competing interests to declare.

\bibliographystyle{unsrtnat}
\bibliography{sample}

\end{document}